\documentclass[printer]{aa}
\usepackage{graphics,longtable}
\usepackage{graphicx}
\usepackage{amsfonts}
\usepackage{txfonts} 
\usepackage{natbib}
\usepackage{amssymb}
\def\kms{km\,s$^{-1}$}

\newcommand{\hi}{\ion{H}{i}}

\begin{document}
\authorrunning {Corti et al.}
\titlerunning{}
\title{GS305+04-26: Revisiting the ISM around the Cen\,OB1 
stellar association }

\author{M. A. Corti\inst{1,2}\thanks{Member of Carrera del Investigador
   Cient\'{\i}fico, CONICET-CCT La Plata, Argentina}, E. M. Arnal\inst{1,2 \star},
 R. B. Orellana\inst{2,3 \star}}
\offprints{Mariela A. Corti,  \email{mariela@fcaglp.unlp.edu.ar}}  

\institute{Instituto Argentino de Radioastronom\'{\i}a (CCT-La Plata, CONICET),
 C.C. No. 5, 1894 Villa Elisa, Argentina
\and 
Facultad de Ciencias Astron\'omicas y Geof\'{\i}sicas,
Universidad Nacional de La Plata, Paseo del Bosque s/n,
1900 La Plata, Argentina
\and 
Instituto de Astrof\'{\i}sica de La Plata (CCT-La Plata, CONICET), Argentina}

\date{Received \today; accepted \ldots }

\abstract{Massive stars deeply modify their surrounding {\rm ISM} via their high throughput of ionizing photons and their strong stellar winds. In this way they may create large expanding structures of neutral gas.}{We study a new large H{\sc i} shell, labelled {\rm GS}305+04-26, and its relationship with the {\rm OB} association Cen\,OB1.} {To carry out this study we have used a multi-wavelenght approach. We analyze neutral hydrogen (H{\sc i}) line data retrieved from the Leiden-Argentina-Bonn ({\rm LAB}) survey, new spectroscopic optical observations obtained at CASLEO, and make use of proper motion databases available via Internet.}
{The analysis of the H{\sc i} data reveals a large expanding structure  {\rm GS}305+04-26 centered at ({\it l,b})=(305$^{\degr}$, +4$^{\degr}$) in the velocity range from -33 to -17 \kms. Based on its central velocity, -26 \kms, and using standard
galactic rotation models, a distance of  2.5$\pm$0.9 kpc is inferred. This structure, elliptical in shape, has major and minor axis of  440 and 270 pc, respectively.  Its expansion velocity, total gaseous mass, and kinetic energy are
$\sim$ 8 \kms, (2.4$\pm$0.5) $\times$ 10$^5$ M$_\odot$, and (1.6$\pm$0.4) $\times$ 10$^{50}$ erg, respectively. Several stars of the {\rm OB}-association Cen\,OB1 are seen projected onto, and within, the
 boundaries of {\rm GS}305+04-26. Based on an analysis of proper motions, new
members of Cen\,OB1 are identified. The mechanical energy injected by these
 stars could have been the origin of this H{\sc i} structure.}{}


\maketitle
\titlerunning{GS305+04-25: Revisiting the ISM around the Cen\,OB1 
stellar association}


\section{Introduction}

Neutral hydrogen emission at $\lambda \sim$ 21-cm shows that the interstellar
 medium ({\rm ISM}) of spiral galaxies have a complex morphology, because
 superimposed on their large scale structure a plethora of different
 features (arcs, bubbles, chimneys, filaments, holes, loops, shells,
 supershells, worms, etc) are observed. Among them, large neutral hydrogen
 shells and their major siblings, the so-called H{\sc i} supershells, are some of the most 
 spectacular phenomena. Though the origin
of the large shells can well be explained by the action of either stellar
 winds or supernova explosions, or most likely, by their combined effects, the
 origin of the H{\sc i}  supershells is not well understood yet. 
The later were 
 discovered by \citet{hei79,hei84}. Large shells of neutral hydrogen and H{\sc i}  supershells
 are usually identified, in a given velocity range, as a minimum in the H{\sc i} emissivity 
distribution surrounded by regions of higher emissivity. 
 In order to power the H{\sc i}  supershells
 an unrealistically large number of massive stars,
ranging from hundreds to thousands, would be required. For these extreme cases,
 alternative processes like gamma-ray bursts \citep{per00} or the infall of
 high
velocity clouds \citep{ten81} have been invoked. Nevertheless, there is a
 handful of the catalogued
 H{\sc i} supershells for which a particular OB-association has been identified as
its likely
 powering source. Among these, \citet{mcc01b} suggest that the
H{\sc i} supershell {\rm GSH}305+01-24 was ''{\it formed from stellar winds in
 the Centaurus OB1 association}''. Among the arguments in favour of such 
interpretation the authors quoted that ''{\it the morphology of the shell seems
 to trace the stellar distribution, suggesting an association}''. Furthermore, 
whilst addressing the issue of whether or not {\rm GSH}305+01-24 could be
 explained as a wind-driven bubble created by the early type stars of Cen\,OB1,
 the Wolf-Rayet star {\rm WR}\,48 ($\theta$\,Mus) was considered to be a member 
of the association.

As part of a long term program aimed at improving the optical data 
(e.g. radial velocity, binarity, spectral types, membership, etc.) of stars 
considered to be members of {\rm OB}-associations located in the Southern
 Hemisphere, new spectroscopic observations were carried out towards most of
 the  early type stars listed by \citet{hum78} as members of Cen\,OB1. 
Furthermore, using Tycho-2  catalogue and applying a new method of proper motion
 analysis recently developed
by \citet{ore10}, a substantial revision of the stars thought to be 
members of Cen\,OB1 is made. 

In the light of the above results it was found that the census of early type
 stars, those that matter from the wind injection energy viewpoint, is
 altered to such an extent that makes worth the effort of reviewing the
 suggested association between Cen\,OB1 and {\rm GSH}305+01-24 claimed 
by \citet{mcc01b}.

\section{ Observational Data}

This research was carried out using both new observational data and
databases publicly available, namely:
\begin{itemize}
\item [{\it a)}] Spectroscopic observations of sixteen (16) stars listed
by \citet{hum78} as members of Cen\,OB1 were obtained at Complejo 
Astron\'omico El Leoncito (CASLEO)\footnote{Operated under agreement between 
CONICET, SeCyT, and the Universities of La Plata, C\'ordoba and San Juan, 
Argentina} during April 2009. The  spectra were obtained using a REOSC Cassegrain
 echelle spectrographs attached to the 2.15-m ''Jorge Sahade'' telescope. The
detector was a TEK CCD (1024 x 1024 pixels) having pixel size of 24$\mu$m.
 A grating of 400 l/mm was used as cross disperser, and the slit width was set 
either to 250$\mu$m or 300$\mu$m. The reciprocal dispersion was 
6.6 \AA/mm. These spectra covered the wavelength range from 3800\AA~to 6500\AA,
and the signal-to-noise ratio is 20 $\leq$ S/N $\leq$ 50.

A He-Ar comparison lamp images were obtained at the same telescope position as
 the stellar images immediately after or before the stellar exposures. Bias
 frames were also obtained every night, as well as spectra of the stars HR 2806
 and HR 7773 as radial velocity standards. The spectra of several spectral
type standards were also obtained along the observing run. All spectra were 
processed and analysed 
using the {\rm IRAF}\footnote{IRAF is distributed by NOAO, operated by AURA, 
Inc. under an agreement with NSF} package.                                                                                         
\item [{\it b)}] Neutral hydrogen line data retrieved from the
Leiden-Argentina-Bonn ({\rm LAB}) survey \citep{kalbe05}. 

\item [{\it c)}] {\rm WR}\,48 proper motion measurements retrieved from the 
{\it Centre de Done\'es Astronomiques de Strasbourg}
 ({\rm CDS}\footnote{http://cdsweb.u-strasbg.fr/}).
\end{itemize}

 \section{Results}

\subsection{Optical spectroscopic}

\begin{table*}[h!]
\begin{center}
\leavevmode
\caption{New spectral types and radial velocity measurements of Cen\,OB1 members (according to \citet{hum78}).}
\label{table:one}
\vskip 0.4cm
\begin{tabular}{|c|c|c|l|c|r|c|l|}
\hline
\hline
 ID & l & b & SpT & V &  V$_{LSR}$ & d  & Commentaries\\
  &($\circ$) & ($\circ$)& & (mag.)& (km s$^{-1}$) & (kpc) & \\
\hline
   HD 110639 & 302.1 & +1.5 & B1 II-III & 8.5 & -52 $\pm$ 03 & 1.8 $\pm$ 0.5 & \\
   HD 111613 & 303.0 & +2.5 & B9-A0 Iab & 5.7 & -27 $\pm$ 03 & 1.5  $\pm$ 0.4 & \\
   HD 111904 & 303.2 & +2.5 & B9 Ia & 5.7 & -25 $\pm$ 03 & 2.6 $\pm$ 0.6 &  NGC 4755\\
   HD 111934 & 303.2 & +2.5 & B2 Ib & 7.0 &  -31 $\pm$ 02 & 2.0 $\pm$ 0.5 &  NGC 4755\\
   HD 111973 & 303.2 & +2.5 &  B5 Ia & 5.9 & -18 $\pm$ 04 & 2.7 $\pm$ 0.6 &  NGC 4755\\
   HD 111990 & 303.2 & +2.5 & B3 Ib & 6.8 & -7 $\pm$ 03 & 1.8 $\pm$ 0.5 & NGC 4755 \\
   HD 112364 & 303.6 & +3.1 & B0.5 Ib-a & 7.4 & -68 $\pm$  12 & 2.8 $\pm$ 0.6 & SB2 \\
   HD 112842 & 304.1 & +2.5 & B3 Ib & 7.0 & -33 $\pm$ 03 & 2.1 $\pm$ 0.5 & \\
   HD 113012 & 304.3 & +2.7 & B0 Ia & 8.1 & -22 $\pm$ 05 & 6.9 $\pm$ 1.6 & \\
   HD 113422 & 304.5 & +1.1 & B1 Ia & 8.2 & -57 $\pm$ 04 & 3.2 $\pm$ 0.7 & \\
   HD 113432 & 304.4 & - 0.7 & B1 Ib & 8.9 & -21 $\pm$  11 & 2.7  $\pm$ 0.6 & \\
   HD 113708 & 304.6 & - 2.4 & B0.2 III & 8.1 & -16 $\pm$ 07 & 3.1 $\pm$ 0.8 & \\ 
   HD 114213 & 305.2 & +1.3 & B1 Ia-b & 8.9 & -2 $\pm$ 06 & 2.2 $\pm$ 0.5 & \\
   HD 114886 & 305.6 & - 0.9 & O9 II-III & 6.9 & -25 $\pm$ 08 & 2.6 $\pm$ 0.6 & SB2 \\
   HD 115363 & 305.9 & - 1.0 & B1 Ia & 7.8 &  -66 $\pm$ 08 & 3.3 $\pm$ 0.9 &  \\
   HD 116119 & 306.7 & +0.6 & B8 Ia & 7.9 & -34 $\pm$ 03 & 4.3 $\pm$  1.1 & \\ 
 \hline
\end{tabular}
\end{center}
\end{table*}

Most of the Cen\,OB1 stars listed by Humphreys (1978) having spectral types
earlier than A2 were observed using the {\rm CASLEO} telescope. They are
listed in Table \ref{table:one}. Using these data and the spectral type catalogue of
 \citet{Wal90} a new spectral classification was derived
for each star (see column 4th in Table \ref{table:one}). The star 
identification is given in the first column, whilst its galactic
coordinates (longitude and latitude) are given in the second and third columns,
 respectively.
Using this new spectral types, the intrinsic (B-V)$_o$ colour index
 and the absolute visual magnitude M$_v$ matching the spectral type were
 obtained from \citet{lando82}.
 With these parameters and adopting a visual extinction A$_v$ = 3.1 E(B-V),
 where E(B-V) is the stellar colour excess, the stellar distance (seventh 
column in Table \ref{table:one}) of each star is derived. The photometric magnitudes 
V and B were corrected from the Tycho-2{\footnote{http://www.rssd.esa.int/index.php?page=Overview\&project=HIPPARCOS} 
system to the Johnson system. The quoted error in the distances stems from an assumed M$_v$ uncertainty of
 0$\cdot$5 magnitudes \citep{wal72}. 

Using the {\rm IRAF} standard reduction package, the stellar heliocentric
radial velocity was derived from a gaussian fit to the He\,I and He\,II lines
present in the stellar spectrum. Later on this heliocentric radial velocity was 
corrected to the Local Standard 
of Rest (LSR) (see sixth column Table \ref{table:two}). Since all program stars but 
two ({\rm HD}\,113432 and {\rm HD}\,114213) 
were observed twice, those stars having highly discordant individual radial velocity 
estimates, were identified as possible multiple (SB2 or triple) systems.
Based on the radial velocity obtained for those stars assumed to be single, 
the mean radial velocity of Cen\,OB1 is -24 $\pm$ 14 \kms. Throughout this
paper, all radial velocities are referred to the Local Standard of Rest ({\rm LSR}).

\begin{figure*}
\centering
\includegraphics[width=12cm]{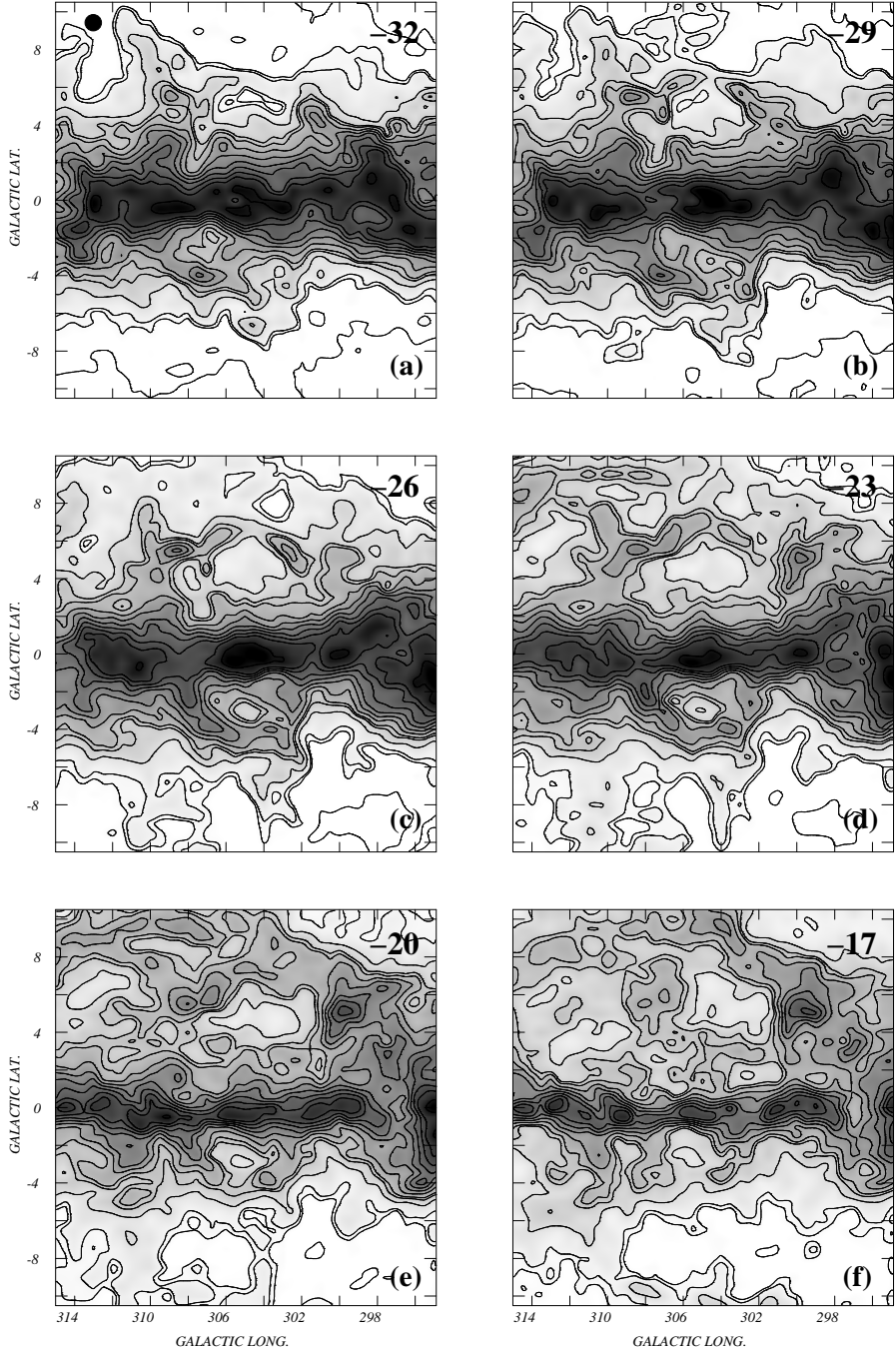}
\caption{Mosaic of H{\sc i} mean brightness temperature in selected
velocity ranges within -33 to -16 \kms. Each image is a mean of three individual images. The
 central velocity of each image is indicated at the inner top right corner.
 The filled circle drawn in the upper left corner of the H{\sc i} image at -32 \kms 
indicates the angular resolution of the H{\sc i} data. 
The lowest and highest contours are  15\,K and 105\,K, respectively, while
the contour spacing is 5\,K until 50\,K and 15 K from there onwards. The dark grey areas represent 
regions with high H{\sc i} emissivity.}

\label{fig:uno}
\end{figure*}

\begin{figure*}
\centering
\includegraphics[width=12cm]{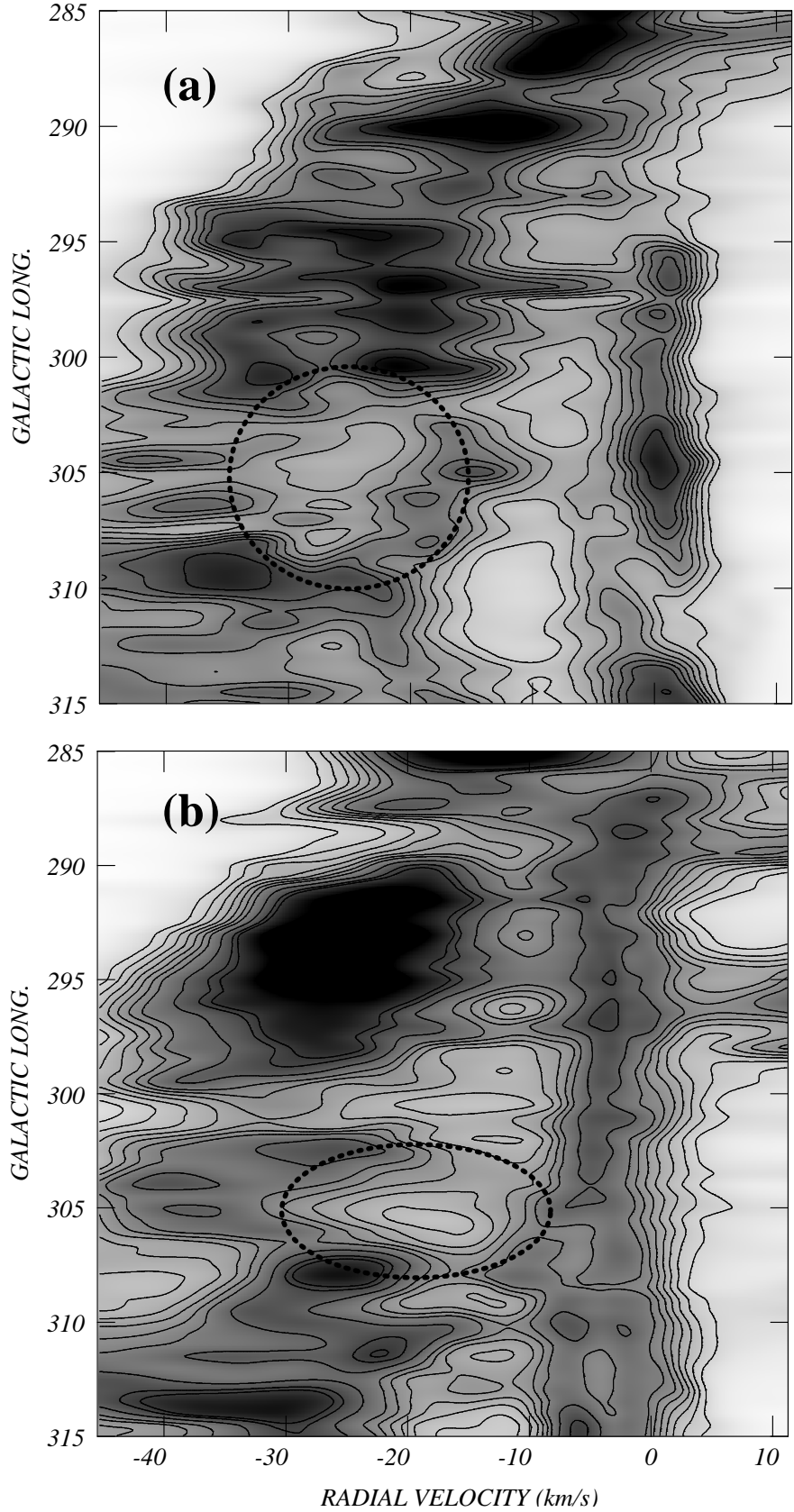}
\caption{Velocity-galactic longitud diagram of
 the mean brightness temperature distribution in two different ranges of
galactic latitude. The upper panel shows the velocity-galactic longitud 
diagram for 3\fdg5 $\leq$ {\it b} $\leq$ 4\fdg0. The dotted circumference
signals the location of Feature {\rm A}. (see text). The lower panel shows the 
same diagram for the galatic latitude strip -3\fdg0 $\leq$ {\it b} 
$\leq$ -2\fdg5. The dotted ellipse marks the location of Feature {\rm B}.}  
\label{fig:dos}
\end{figure*}

\begin{figure*}
\centering
\includegraphics[angle=-90.,width=12cm]{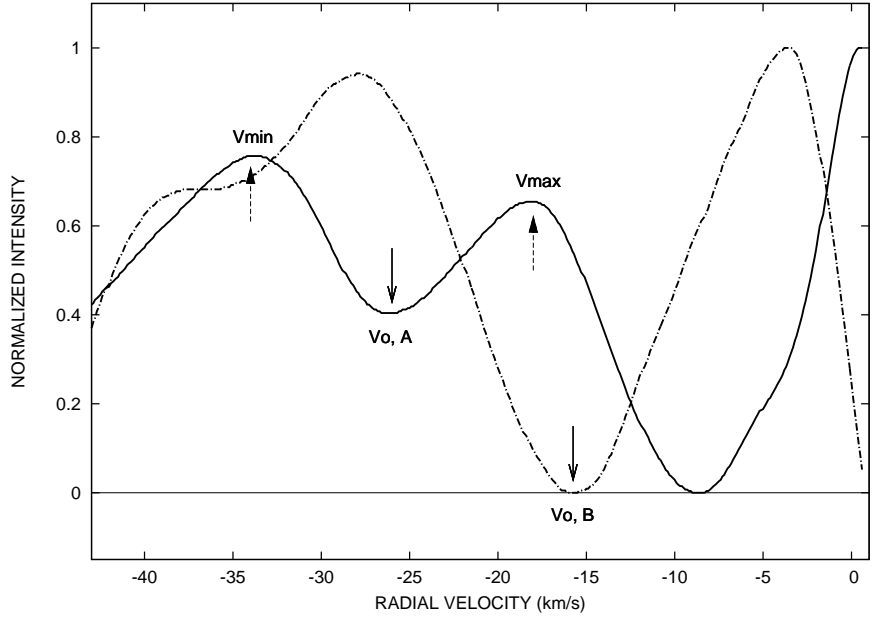}
\caption{Normalized mean H{\sc i} brightness temperature profiles
for the range  298$^{\circ}$ $\leq$ {\it l} $\leq$ 312$^{\circ}$  (Feature {\rm A}, 
continuos line) and between  300$^{\circ}$ $\leq$ {\it l} $\leq$ 310$^{\circ}$  
(Feature {\rm B}, dash-dotted line profile) . The arrows mark the mean radial velocity for
Feature {\rm A} (V$_{o,A}$) and Feature {\rm B} (V$_{o,B}$), respectively.}
\label{fig:3n}
\end{figure*}

\begin{figure*}
\centering
\includegraphics[width=10cm]{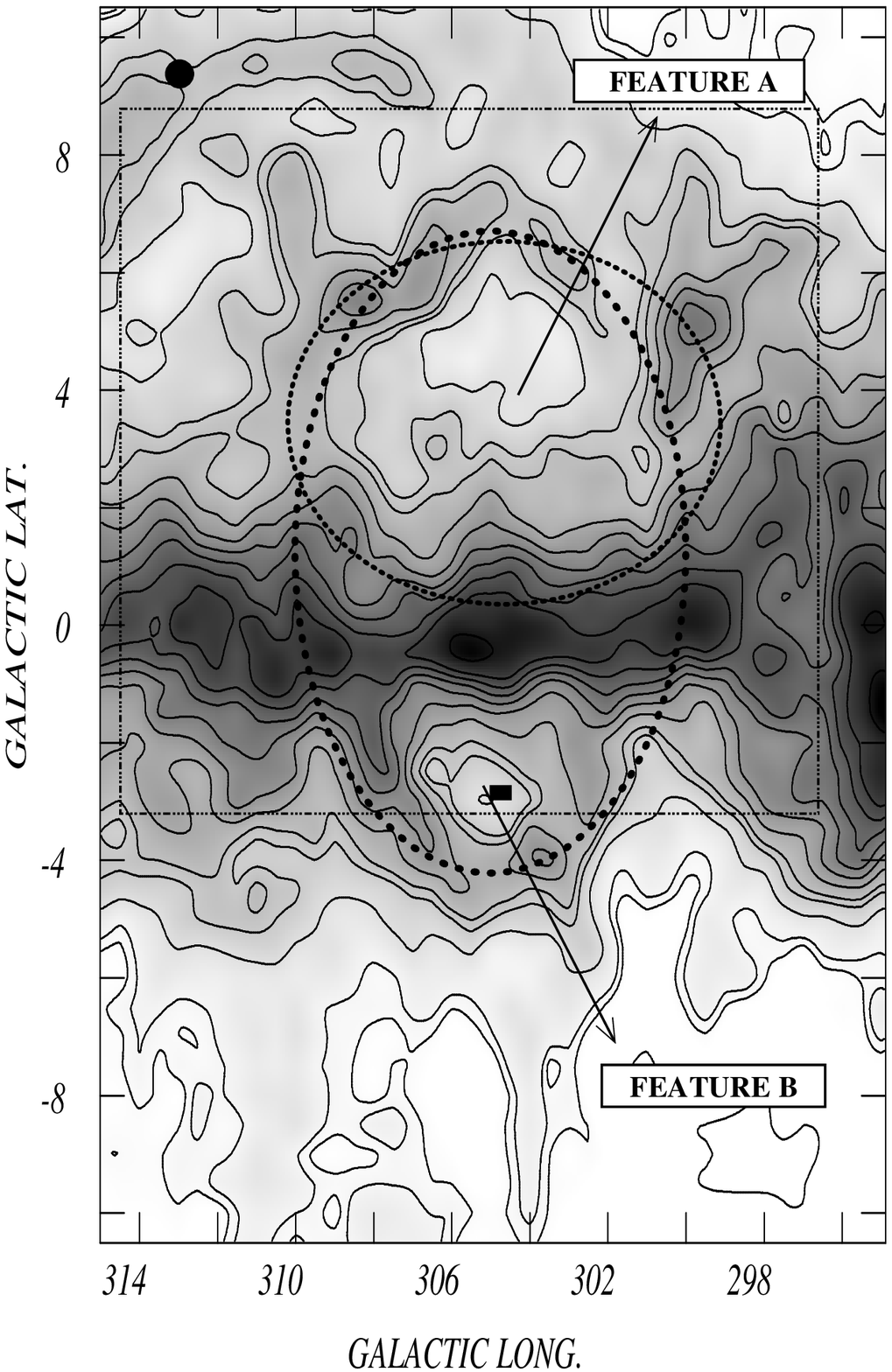}
\caption{Grey-scale representation of the mean H{\sc i} brightness temperature
 in the 
velocity range -26 to -19 \kms. The light dot ellipse represents the least
square fit to the H{\sc i} peaks defining {\rm GS}305+04-26. The location of {\rm GSH}305+01-24  \citep{mcc01b} 
is depicted by a dark dot ellipse. The dash-dotted rectangle delimits the 
region shown in Fig.\ref{fig:cuatro}. The small black rectangle 
shows the position of the star {\rm WR}\,48. The angular resolution of these 
data is given by the filled dot drawn in the upper left corner.}
\label{fig:tres}
\end{figure*}

\subsection{Proper motion analysis}
\subsubsection{{\rm WR}\,48}

 Large scale catalogues (Hipparcos 1997; 
Tycho2 2000; PPMX 2008; ASCC2.5 2009) 
quote quite different, and a first glance even contradictories, proper motions
 for the Wolf-Rayet {\rm WR}\,48 ($\equiv${\rm HD}113904$\equiv$$\Theta$Mus). 
 All proper motions in this
paper will be expressed in units of milliarcseconds per year (mas/yr). From a
 thorough analysis
of the proper motions provided for this star by the different catalogues,
 those given
 by the ASCC2.5 catalogue ($\mu_{\alpha}cos{\delta}$, $\mu_{\delta}$)=
(-4.42$\pm$0.79, 2.53 $\pm$0.95) \citep{khar01} are found to be the most
 reliable ones and will be adopted in this paper.
 This conclusion is based on the following considerations: 

\begin{itemize}
\item[{\it a)}] It is a well known fact that due to the short time interval covered by the Hipparcos 
mission (less than four years), quite often the proper motions provided by this catalogue are not as good as their 
formal standard errors seems to indicate. This situation is even worse for double or multiple stars 
\citep{urb00,von04}, 
because in these cases the quoted proper motions may reflect the {\it instantaneous} proper motion 
rather than the {\it mean} value needed for long term extrapolation. The Hipparcos catalogue quotes
the components of the proper motion 
($\mu_{\alpha}cos{\delta}$, $\mu_{\delta}$)=(-5.35$\pm$0.59, -2.23$\pm$0.68) 
 \citep{esa97}.

\item[{\it b)}] The reliability of the components of the proper motions of
{\rm WR}\,48 ($\mu_{\alpha}cos{\delta}$, $\mu_{\delta}$)=(-1.9$\pm$1.3, 
19.0$\pm$1.3) \citep{hog00} quoted by the Tycho-2 catalogue,
could be evaluated in the light of 
the so-called ''{\it goodness of the fit}'', that is given by a quality flag parameter
(see columns 18 and 19 of Tycho-2 catalogue\footnote{
http://webviz.u-strasbg.fr/viz-bin/VizieR?-source=I\%2F259}). ``{\it This goodness of fit is 
the ratio of the scatter-based and the model-based error}'',
 and may get values ranging from 0 up to 9. A larger value should be taken as 
a warning on the proper motion quality. For {\rm WR}\,48 the proper motion in declination has 
a quality flag of 7.5.

\item[{\it c)}] Proper motions retrieved from the Position and Proper Motion Extended catalogue
({\rm PPMX}) \citep{ros08}  having a problem with the least square fit used to derive the proper 
motion of a given star were signaled with a flag ''{\rm P}''. Those quoted for {\rm WR}\,48
($\mu_{\alpha}cos{\delta}$, $\mu_{\delta}$)=(-0.3$\pm$1.7, 16.6$\pm$1.8) have such a flag.

\end{itemize}

\subsubsection {Cen\,OB1}

The stellar association Cen\,OB1 was first reckoned as such based on 
spectrophotometric study of \citet{hum78}. A total of 29 stars were listed as
 members of this association (see Table 8 of \citet{hum78}). The distance
 modulus derived by \citet{hum78} for this association is 11.9 magnitude
 ($\sim$ 2.4 kpc). Later on, based on a photometric work of \citet{hum84}, the
 members of Cen\,OB1 was increased to 106 stars. 

Since long time ago stellar proper motions have been used to established the
 likelihood of membership of a given star to a given stellar grouping such as an
open cluster or an OB-association \citep[and references therein]{dia02,deze99}. 
Nowdays, the availability of
proper motion catalogues containing thousands of measurements of stars (e.g. 
Hipparcos 1997; Tycho2 2000; PPMX 2008; ASCC2.5 2009)
 distributed all over the sky, offers the chance of routinely using
proper motions-based techniques in order to improve our knowledge on the
number of stars that may be related to an already known stellar grouping.

Since according to \citet{mcc01b} {\rm GS}305+01-24 was {\it ''
formed by the stellar winds in the Centaurus OB1 association ''}, it is 
worth making an identification as likely as possible of its members.

To identify the members of Cen\,OB1 based on the analysis of existing large 
scale proper motion catalogues, a two steps procedure was followed. Firstly,
the technique of \citet{ore10} was applied to the Hipparcos Catalogue (that
is complete down to a visual ({\rm V}) magnitude ranging from 7.3 to 9.0 
(depending on spectral type), on a circular area of radius 7$^{\degr}$ centered
 at  ({\it l,b})=(305$^{\degr}$,+0$^{\degr}$). The reason for selecting 
such a
 large search radius is twofold, namely: {\it i)} To be able to search for stars
 candidate to be members of Cen\,OB1 on an angular area larger than the one 
covered by the stars quoted by \citet{hum78} as members of the association;
{\it ii)} To include all stars located within a solid angle comparable to, or
 slightly larger than the region covered by {\rm GS}305+01-24 \citep{mcc01b}.
Having in mind that the short time interval covered by the Hipparcos mission
conspires against the accuracy of its quoted proper motions, in a second step 
 the same technique was applied 
to an square area of 7$^{\degr}$ in size of the Tycho-2 catalogue. This area is
also centered at ({\it l,b})=(305$^{\degr}$,+0$^{\degr}$). Bearing in mind 
 that the Tycho-2 catalogue is 
90$\%$ complete down to a visual ({\rm V}) magnitude V=11.5} we should be able
 to identify all
supergiant stars, regardless of their spectral type, that may belong to 
Cen\,OB1, and all main sequence and giant stars whose spectral types are
earlier than {\rm B}\,3 V and {\rm B}\,6 III, respectively. In this way,
 105 stars were identified as possible members of Cen OB1.

Restricting the above sample only to those stars having
both a spectral type classification and photometric data, and retaining from
this sub-sample only those stars whose distance lie within $\pm$1$\sigma$ of 
the mean distance, a total of 54 stars were identified as likely candidates
to be members of Cen\,OB1. These stars are listed in Table \ref{table:two}.
The mean spectrophotometric distance of this grouping turns out to be 
2.6$\pm$0.4 kpc, 
in good agreement with previous Cen\,OB1 distance estimates 
\citep{hum78,hum84,kalt94}. Comparing the list of 106 stars of
 \citet{hum84}
 with our listing of 54 stars (see Table \ref{table:two}), it is found
 that only $\sim$24\% of the stars are common to both sets. In the same 
 sense, only 10 out of the 22 stars originally listed by \citet{hum78} as members 
 of Cen\,OB1 are present in our sample.

To be considered
as a candidate star to be member of Cen\,OB1, a given star must have a
 membership probability (P$_a$) higher than or equal to 0.5 according to
 the Bayesian criterion. This means that if $\Phi_a$ represents the spatial
 function distribution of all stars in the region belonging to the association
 and 
$\Phi_f$ represents the spatial distribution of all non-member stars present in 
the same region (see Sect. 4.1 of \citet{ore10}), then

\begin{equation}
 P_a=\frac{\Phi_a}{\Phi_f+\Phi_a}
\end{equation}

The components of the mean proper motion of Cen\,OB1 stars are 
 ($\mu_{\alpha}cos{\delta}$, $\mu_{\delta}$)=(-4.65$\pm$0.15, -0.92$\pm$0.12).

A comparison of this proper motion with the one adopted for {\rm WR}\,48, 
indicates that whilst the later is moving from lower galactic latitudes 
towards the galactic plane, the association Cen\,OB1, as a whole,  is moving in 
the other way.
Therefore, {\rm WR}\,48 is unlikely to belong to this association.

\subsection{One single large H{\sc i} structure ?}

 Relying on the central radial velocity (V$_o$= -24 \kms) and the
expansion velocity (V$_{exp}\sim$ 7 \kms) of {\rm GSH}305+01-24 derived by \citet{mcc01b}, 
the H{\sc i} brightness temperature distribution covering the velocity
 range from -40 to -10 \kms was re-analised within the region delimited by
290$^{\degr}$ $\leq$ {\it l} $\leq$ 320$^{\degr}$ and 
-10$^{\degr}$ $\leq$ {\it b} $\leq$ +10$^{\degr}$ using the {\rm LAB}
H{\sc i} survey \citet{kalbe05}. Though this survey has a lower
angular resolution ({\rm HPBW}=34\arcmin) than other available databases 
(e.g. \citet{kalbe05} , {\rm HPBW}=14\farcm4), the difference in angular resolution plays
only a minor role when the object under study, like the one we are dealing 
with, has angular dimensions of several degrees. The main observational
characteristics of this distribution are shown in Fig.\ref{fig:uno}, where a
 mosaic of six H{\sc i} images spanning the velocity from -32 to -17 \kms is
 shown.

Before describing the H{\sc i} images,  a word of caution is in place. 
Considering that towards the inner part of the Galaxy the radial 
velocity-to-distance transformation is double-valued, the straightforward
 interpretation that the H{\sc i} minima observed, for example, in 
Fig. \ref{fig:uno}(c) represent {\it different} structures may not be 
correct, because in a given radial velocity range, H{\sc i} emission arises 
from two different locations along the line of sight. Therefore, a large 
single H{\sc i}
minima extending both above and below the galactic plane will {\it always} 
appear bisected at lower galactic latitudes by a strong band of galactic 
H{\sc i} emission, giving the impression of two different minima when in 
reality is a single structure. Having clarify this point, we shall proceed to 
briefly described, only from a morphological point of view, the main characteristics 
of the H{\sc i} emission observed along the velocity range -32 to -17 \kms.
The images at
-26 \kms 
(Fig.\ref{fig:uno}(c)), -23 \kms (Fig.\ref{fig:uno}(d)), and -20 \kms 
(Fig.\ref{fig:uno}(e)) the H{\sc i} distribution shows the presence of two well 
developed H{\sc i} minima in the brightness temperature distribution. The one
located at positive galactic latitudes (labelled Feature {\rm A} from here
 onwards) roughly centred at ({\it l,b})=(305\degr, +4\degr) is ovoidal in 
shape ($\bigtriangleup${\it l}, $\bigtriangleup${\it b}) $\sim$ 
(10\degr, 6\degr) and has its major axis almost paralell to the galactic plane.
The minimum located below the galactic plane (Feature B from here onwards), also
elliptical in shape, is centred at ({\it l,b})=(305\degr, -3\degr) and has its
 major axis along a position angle of $\sim$ 60\degr. Position angles are measured
counterclockwise from north galactic pole direction. At more negative
radial velocities than those mentioned above (see Fig.\ref{fig:uno}) 
Feature {\rm A} remains visible till -32/-30 \kms, where it begins to become
 undetectable
against the overall galactic H{\sc i} emission at those velocities. 
On the other hand, Feature {\rm B} is hardly seen at -32 \kms (Fig.\ref{fig:uno}(a)).
The noticeable H{\sc i} minimum seen at ({\it l,b})=(307\degr, -2\fdg5)
should not be mistaken with Feature {\rm B}, because the former, first
detected at -29/-28 \kms, is located closer to the galactic plane and towards
increasing galactic longitudes than Feature {\rm B}.
At radial velocities more positive than -23 \kms, though both H{\sc i} minima
 are easily identifiable in the H{\sc i} distribution at -20 \kms, Feature 
{\rm A} begins to loose identity at -17 \kms, whilst at the same velocity 
 Feature 
{\rm B} remains clearly visible. Therefore, from Fig.\ref{fig:uno} it appears 
that Feature {\rm A} and Feature {\rm B} may not be detectable
 along the {\it same velocity} 
interval. In order to further analyse this point, radial velocity versus galactic
 longitude images, at specific intervals of galactic latitude, were constructed.
 In 
Fig.\ref{fig:dos} two such images are shown. The upper panel depicts the mean H{\sc i}
 emission distribution at 3\fdg5 $\leq$ {\it b} $\leq$ 4\fdg0, whilst the
lower one corresponds to the galatic latitude interval -3\fdg0 $\leq$
 {\it b} $\leq$ -2\fdg5. In both panels the local H{\sc i} emission is depicted  by the 
 almost straight ridge of emission seen at both slightly positive 
({\it upper panel}) and slightly negative velocities ({\it lower panel}). The
minimum of H{\sc i} emission corresponding to either Feature {\rm A} or Feature {\rm B} 
are encircled in both panels by thick-dashed figures 
to help the 
reader to identify them. Admittedly, the above argument is only a
qualitative one. In Fig. \ref{fig:3n} the normalized mean brightness
temperature as a function of radial velocity, as derived from 
Fig. \ref{fig:dos} for different galactic longitude ranges, are shown. 
In Fig. \ref{fig:3n} the
continuos line shows the normalized mean brightness temperature for the 
range  298$^{\circ}$ $\leq$ {\it l} $\leq$ 312$^{\circ}$ of Fig. \ref{fig:dos}a), whilst
the dot-dashed line shows the behaviour observed for Feature {\rm B} in
Fig. \ref{fig:dos}b). In this case the interval used to construct the
normalized H{\sc i} profile is 300$^{\circ}$ $\leq$ {\it l} $\leq$ 310$^{\circ}$ . 
The normalized emission was computed by substracting from each point the
 minimum value of the corresponding scan (19\,{\rm K} and {\rm31}\,K for
 Features {\rm A}
 and Feature {\rm B}, respectively) and then dividing by the corresponding
 maximum (27.2\,{\rm K} and 21.1\,{\rm K} for Features {\rm A} and {\rm B}, 
respectively). Clearly enough, Fig. \ref{fig:3n} shows
that the mean radial velocity and the {\rm FWHM} of both features, namely
V$_{o,A}$=-26 \kms and 9 \kms, respectively, for Feature {\rm A}, and 
V$_{o,B}$=-16 \kms and 14 \kms, respectively, are different. This is not the expected behaviour 
in case {\it both}
features \underline{\it were} part of a single large H{\sc i} void, because
in this case one would expect that  the mean radial velocity of both
features ({\rm A} and {\rm B}) were, within errors, similar and the 
{\rm FWHM} of Feature {\rm B}  were smaller than the corresponding to 
Feature {\rm A}. 
 Mean radial velocities and {\rm FWHM} were derived
from a Gaussian fitting and both are accurate to within $\pm$ 1 \kms.

Using the linear fit and the power law fit of the galactic rotation model of 
\citet{bli89} the kinematic distances of both  H{\sc i} structures turned out 
to be 2.5$\pm$0.9 kpc (Feature {\rm A}) and 1.7$\pm$0.7 
kpc (Feature {\rm B}), respectively. The quoted distances are a mean weighted 
value 
of the individual distances provided by the different models. The weight of the 
individual distance determinations stems from the assumption of an uncertainty 
of $\pm$ 8 \kms due to non-circular motions \citep{bur88}. 

  Therefore, under the assumption that both shells have a
 line-of-sight dimension comparable to those observed in the plane of the sky, and
exclusively relying on the derived mean distances for both H{\sc i} features,
the line of sight width of both shells would be insufficient to overlap them in 
space, even within the large errors quoted for the individual distances of
both features.

Based on above, it is believed that Feature {\rm A} and Feature {\rm B} are 
{\it unrelated} H{\sc i} structures, and not different manifestations of the same feature  
({\rm GSH}305+01-24) as put forward by \citet{mcc01b}. In this
context, it is worth mentioning that Feature {\rm B} was interpreted \citep{cap84} as
 being an H{\sc i} feature exclusively associated with the Wolf-Rayet star {\rm WR}\,48 
 ($\equiv\Theta$Mus), that is not a member of Cen\,OB1.

The mean brightness temperature distribution in the range -26 to -19 \kms is shown 
in Fig.\ref{fig:tres}. The shell {\rm GSH}305+01-24 reported by \citet{mcc01b} is shown 
as a thick dark dotted ellipse, whilst the approximate boundaries of Feature
 {\rm A} 
is depicted by a light dotted line. Following the standard nomenclature,
 Feature {\rm A} from here onwards
will be referred to as {\rm GS}305+04-26. The high latitude border of this
 structure has a mean brightness 
temperature of $\sim$20\,K above the surrounding background galactic emission, a fact 
that favours its clear identification. Unfortunately, the low latitude 
extreme of {\rm GS}305+04-26 is not detected because is projected onto the
 strong (T$_b\sim$ 90\,K)
emission arising from the H{\sc i} located at the far kinematic distance
 ($\sim$\,7.5 kpc) along this line of sight, making the effort of 
 attempting to disentangle
it from the overall galactic H{\sc i} emission a pointless task. The physical
 parameters of {\rm GS}305+04-26 are derived from a least
 square fit of an ellipse to the H{\sc i} peaks defining the shape of 
{\rm GS}305+04-26 above {\it b} = 2$^{\degr}$. From this fit, shown in
 Fig.\ref{fig:tres} 
by the a thin dotted line, the supershell center is 
({\it l,b})=(305$^{\degr}$, +4$^{\degr}$) and the
  major and minor angular semi-axis of {\rm GS}305+04-26 are
 $\sim$ 5$^{\degr}$ and 3$^{\degr}$, respectively. 
The fitting errors of the supershell centre are $\pm$ 0\fdg3. At the
distance of {\rm GS}305+04-26 these angular dimensions are equivalent to 
 220$\pm$70 (R$_{max}$) and 135$\pm$40 (R$_{min}$) pc,
 respectively.

Under the assumption of an optically thin H{\sc i} emission, the total H{\sc i}
 mass of a 
structure located at a distance {\it d} (kpc) that subtends a solid angle $\Omega$ 
(squared arc min) is given by

\begin{equation}
 M_{HI}=0.0013\,d^2\,\bigtriangleup V\,\bigtriangleup T_B\,\Omega\,(M\odot)
\end{equation}

 where $\bigtriangleup$V is the velocity interval covered by the structure,
 expressed in \kms, and $\bigtriangleup$T$_B$ (K) is the mean brightness 
temperature of the structure defined as 
$\bigtriangleup$T$_B$=$\mid$ T$_{sh}$ - T$_{bg}$ $\mid$, where T$_{sh}$
refers to the mean shell H{\sc i} brightness temperature, and T$_{bg}$ is the
brightness temperature defining the outer level of the H{\sc i} structure.
In this way, the neutral hydrogen mass of {\rm GS}305+04-26 is of the order
 of  M$_{\rm HI}$ = (1.8$\pm$0.4) $\times$ 10$^5$  M$_{\odot}$. Assuming a
helium abundance of 34\% by mass, the {\it total} gaseous mass of
 {\rm GS}305+04-26 is about M$_t$ = ( 2.4$\pm$0.5) $\times$ 10$^5$ M$_{\odot}$.

The expansion velocity plays a key role (along with the shell's total
mass) at the time of evaluating the shell's total kinetic energy. One way of
determining this parameter stems from the use of position-position H{\sc i}
images (e.g. those shown in Fig.\ref{fig:uno}) of the feature under study. 
In this
case the expansion velocity (V$_{exp}$) is determined making use of the
maximum approaching (V$_{max}$) and maximum receding (V$_{min}$) radial 
velocities of those H{\sc i} features identified as belonging to the expanding 
structure. Applying this procedure, an expansion velocity 
\-V$_{exp}$= 0.5$\times$(V$_{max}$-V$_{min}$)
 of $\sim$6 \kms is derived. Generally speaking, V$_{exp}$ determined in
this way is a lower limit to the real expansion
velocity, because 'extreme'-velocity gas associated with the expanding H{\sc i}
feature could be missed owing to confusion effects with the overall 
galactic H{\sc i} emission. Another way of determining V$_{exp}$ is
based on the use of velocity-position (either galactic latitude or galactic
 longitude) diagrams. To evaluate V$_{exp}$ the normalized H{\sc i} 
brightness temperature profile shown in Fig. \ref{fig:3n} (continuous line) is
 used. By making a Gaussian fit to the H{\sc i} emission peaks, V$_{max}$ and 
V$_{min}$ can be derived along with V$_{o,A}$ that corresponds to the velocity 
between V$_{max}$ and V$_{min}$ where the minimum value of the H{\sc i} emission
is observed. We derived V$_{o,A}$= $-$26$\pm$ 1 \kms.
The approaching gas is observed at V$_{min}$=$-$34 \kms and
 the receding gas is observed at V$_{max}$=$-$17 \kms. Therefore, the velocity
 extent of the observed \hi\, emission associated with {\rm GS}305+04-26 is 
consistent with an 
expansion velocity of V$_{exp}$ $\simeq$ 8 $\pm$ 1 \kms. Adopting this value for V$_{exp}$
and using the total mass calculated for {\rm GS}\,305+04-26, the kinetic energy 
(E$_k$=0.5\,M\,V$_{exp}^2$) of the expanding structure is E$_k\sim$ 
(1.7 $\pm$ 0.4)$\times$10$^{50}$ erg.

The dynamic  (t$_{dyn}$) age of {\rm GS}305+04-26  was derived adopting
 the \citet{wea77}
analytic solutions for a thin expanding shell created by a continuos injection
of mechanical energy. Under this assumption the dynamic age is given by

\begin{equation}
\label{eq:tdy}
 t_{dyn}= 0.55 \times 10^6 R_s / V_{exp} (yr)
\end{equation} 

where {\rm R$_s$} is the radius of the structure, expressed in units 
of pc. The
expansion velocity is given in units of \kms and the constant in Eq. 
\ref{eq:tdy} represents the mean value between the energy and momentum
 conserving cases. Adopting for R$_s$=$\sqrt{(\mathrm{R}_{max} \times \mathrm{R}_{min})}$, 
the harmonic mean of the major and minor semi-axes,
and using an expansion velocity of 8\,\kms, a t$_{dyn}$$\sim$10$^7$ yr 
is derived.

\section{Discussion}
\subsection{ Origin of {\rm GS}\,305+04-26.}

Could {\rm GS}\,305+04-26 have been created by the stellar winds and supernova
 explosions of the most massive members of Cen\,OB1?. Under the assumption that 
all stars listed in Table \ref{table:two} are actually members of Cen\,OB1,
and using a very simple geometrical model, we shall attempt to answer the
previous question. In our model we shall also assume that all the stars
belonging to Cen\,OB1 were at the beginning closely packed at the centroid of 
{\rm GS}\,305+04-26.

The total mechanical energy (E$_w$) injected by a star along its lifetime is
given by E$_W$=L$_w$$\tau$, where L$_w$ is the stellar mechanical
 luminosity (L$_w$= 0.5\.Mv$_\infty^2$) and $\tau$ is the 
age of the star. Under the assumption of solar metallicity, the stellar mass
 loss rate (\.M) and the wind terminal velocity (v$_\infty$), were derived 
from \citet{jag88},

\begin{equation}\label{eq:mass-loss}
log\, \dot{M} = 1.769 log \, L/L_\odot - 1.676 log \, T_{eff} - 8.158
\end{equation} 

\begin{equation}\label{eq:wind}
 log\, v_\infty = 1.23 - 0.3 log\, L/L_\odot + 0.55 log\, M/M_\odot + 0.64 log\, T_{eff}
\end{equation}

where {\rm L} is the stellar luminosity expressed in solar luminosity units,
and T$_{eff}$ is the effective stellar temperature expressed in K.

 In Eq. (\ref{eq:wind}), {\rm M} indicates the mass of the star (in solar 
mass units). The value of v$_\infty$ agree with observations to within 20$\%$.
 For a given spectral type, the basic stellar parameters ({\rm L, M} and
 {\rm T$_{eff}$}) were adopted from  the values given by \citet{lando82}.

In order to provide an estimate for the stellar mass, using both the spectral
 type given in column fourth of Table \ref{table:two}, and the relationship
between stellar mass and effective temperature given by \citet{lando82}, a mass
estimate (eight column of Table \ref{table:two}) was derived for every star.
Following \citet{scha92} the age ($\tau$) is derived. For those stars
 belonging to the main sequence, an age 
estimate was drawn under the assumption that they are in the nuclear hydrogen
 burning phase, whilst for the evolved stars (those having luminosity classes
 other than {\rm V}) their ages were determined after adding up together the time
 spent in the hydrogen, helium and carbon nuclear burning phases. The stellar 
ages are given in column nine of Table \ref{table:two}. Following the above
procedure, late B--type stars with uncertain spectral classification 
({\rm HD}\,109253, {\rm HD}\,110912, {\rm HD}\,111121, and {\rm HD}\,116864)
 have masses, and hence stellar ages, quite uncertain. Consequently, neither
{\rm M} nor $\tau$ was derived for these four stars.
The total mechanical energy injected by the stars of Cen\,OB1 is the sum
of the contributions of all stars along their lifetime. Based on above, the
total mechanical energy input  by the stars of
Cen\,OB1 into its local {\rm ISM} is ( 2.7$\pm$0.1) $\times$ 10$^{51}$ erg.
Using  \citet{jag88} to estimate the wind terminal velocity, and the 
parameterization of the mass-loss rate of \citet{lam93},

\begin{equation}\label{eq:lamers}
log\, \dot{M} = 1.738 log \, L/L_\odot - 1.352 log \, T_{eff} - 9.547
\end{equation} 

 results
in a total mechanical energy input lower by almost 30\%. Equation (\ref{eq:lamers})
is the average relation for \.M in a sample of Galatic O stars studied by
\citet{lam93}. According to these authors the standard deviation of
individual mass-loss rates is $\sigma$=0.23.

Though theoretical models predict that 20\% of the wind injected energy is 
converted into mechanical energy of an expanding shell \citep{wea77},
from an observational view point the energy conversion efficiency seems to be
as low as 2--5 \% \citep{cap03}.

Additional sources of energy input may have been present within the boundaries 
of {\rm GS}\,305+04-26. It is worth pointing out that supernovae explosions 
({\rm SN}$_e$) may have
taken place among the most massive members of Cen\,OB1. Though there are no
supernova remnants catalogued within the solid angle covered by 
{\rm GS}\,305+04-26, three pulsars located within the boundaries of this shell
indicate that {\rm SN}$_e$ took placed in the past. The pulsars are
PSR J1253-5820 (d = 2.2 kpc, ({\it l, b}) = (303\fdg2, +4\fdg5)), 
PSR J1334-5839 (d = 2.4 kpc, ({\it l, b}) = (308\fdg5, +3\fdg7)), and 
PSR J1254-6150 (d = 2.24 kpc, ({\it l, b}) = (303\fdg3, +1\fdg02)).
The characteristic age of the pulsars ($\tau_c$) are $\sim$2$\times$10$^6$ yr
(PSR J1253-5820), $\sim$5$\times$10$^6$ yr (PSR J1254-6150), and  
$\sim$ 8$\times$10$^7$ yr (PSR J1334-5839), respectively \citep{man05}. The
distance of PSR J1253-5820 has been taken from \citet{nou08}, whilst the
distance of the other two is given by \citet{man05}.  

From Table \ref{table:two} there is evidence of continuous star formation, 
because the stellar ages $\tau$ covered the range from $\sim$ 5 to 24 Myr,
 with a mean value of 12.6 $\pm$ 5 Myr. Bearing in mind the fact of a continuous
stars formation, we shall adopt as the 
age of the association the age of the oldest stars likely to be members
of Cen\,OB1. Hence, Cen\,OB1 would be $\sim$2$\times$10$^7$ years old.
The ages of PSR J1253-5820 and PSR J1254-6150 are
consistent with the assumption that they could have been born in {\rm SN}$_e$
of stars members of Cen\,OB1. On the other hand, PSR J1334-5839 has an age
almost an order of magnitude greater that the age of Cen\,OB1. Bearing this
in mind, the massive progenitor star of this pulsar very likely was not a 
member of Cen\,OB1 and played no role in the formation of  {\rm GS}305+04-26.

Under the assumption that the progenitors of both pulsars were members of
Cen\,OB1, they may have contributed to the formation of {\rm GS}305+04-26 in 
two ways. First, by injecting mechanical energy along their stellar lifetime, 
and
secondly by their final {\rm SN}$_e$. A lower limit to the mechanical energy
input by the stellar progenitors of the pulsars prior to their explosion as 
{\rm SN}$_e$, could be derived assuming that each stellar progenitor spent
a time as a ''stellar" object, equal to the age of the oldest star likely to be member 
of Cen\,OB1 ({\rm HD}\,109937)  minus the pulsar's age. In this way,
the progenitors of each pulsars could have been main sequence stars in the
range from 10 to 12 M$_\odot$. Using Eqs. \ref{eq:mass-loss} and \ref{eq:wind},
 both progenitors stars could have injected an extra mechanical energy of
 $\sim$ 2.4$\times$10$^{51}$ erg during their lives as stars.

The effect of supernovae on wind-driven bubble evolution may be similar to that
of stellar winds acting as an input energy source \citep{mccr87}.
 These authors show that assuming a standard initial mass function for a 
stellar association having N$_{\ast}$ stars with mass greater that 7 M$_{\odot}$, 
the mean energy input by  {\rm SN}$_e$ is given by
E$_{SN}$$\sim$ 2.0$\times$10$^{49}$\,N$_{\ast}$E$_{51}$\,$\tau_{OB}$ erg, where
E$_{51}$ is the supernova explosion expressed in units of 10$^{51}$ erg and
$\tau_{OB}$ is the association's age given in units of 10$^6$ yr. 
Inserting N$_{\ast}$=2 and $\tau_{OB}$$\simeq$ 24 Myr, we obtain
 E$_{SN}$$\sim$10$^{51}$ erg. Certainly, the assumption that only two SN$_e$ 
 (those that originated the pulsars we are observing nowdays) took place among the 
 stars of Cen\,OB1 must be viewed as a gross simplification because in
the past more stars of Cen\,OB1 could have gone through the SN$_e$ phase 
without leaving observable evidence that those events did happen. If the
 end stellar product of those hypothetical explosions were pulsars, they could
 be there and at the same time remain undetected due to their highly
 anisotropic emission. Therefore, by assuming that only two SN$_e$ took place,
 a lower limit is provided to the total energy input in this way to the
 {\rm ISM} local to Cen\,OB1.

Therefore, adding up the energy injected into the {\rm ISM} by the stellar
 winds of the stars of Cen\,OB1 and the progenitor stars of both pulsars, and 
the energy injected by both {\rm SN}$_e$, a total of 
(3.6$\pm$0.1)$\times$10$^{51}$ erg may have been injected into the {\rm ISM}
 by the stars of Cen\,OB1 along their lives. In this case the kinetic energy of
 {\rm GS}305+04-26 would imply an energy conversion efficiency of $\sim$3$\%$\,
in very good agreement with the results of \citet{cap03}.

\subsection{Explaining the off-centre location of Cen\,OB1}

\begin{figure*}
\centering
\includegraphics[width=12cm]{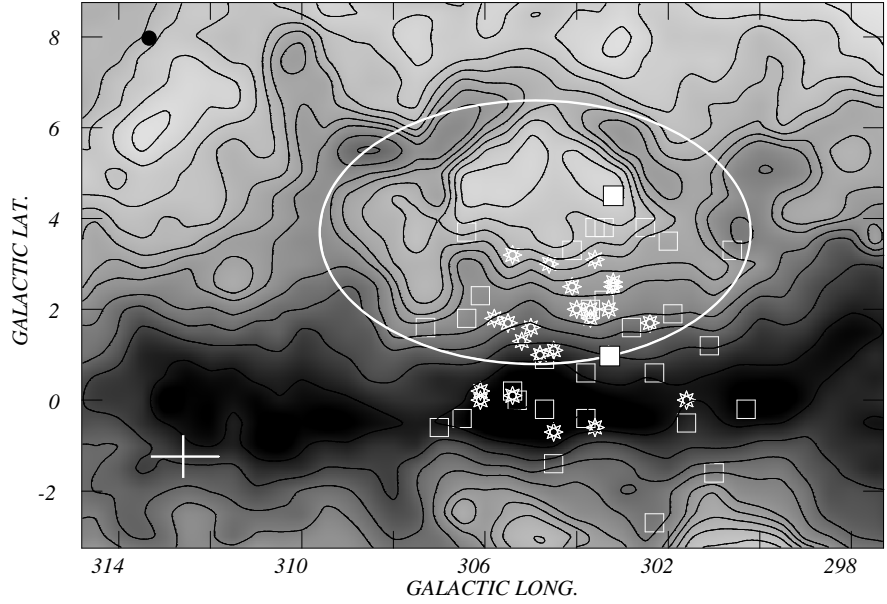}
\caption{Blow up of the region delimited by a dash-dot rectangle in the Fig.
\ref{fig:tres}. The angular resolution of these 
data is given by the filled dot drawn in the upper left corner of the image.
The white thin line ellipse
represents to {\rm GS}305+04-26. The white boxes mark the location of the pulsars. The star symbol indicate the location of Cen\,OB1 members identified
as such by both \citet{hum84} and Corti \& Orellana (in preparation), whilst
the unfilled white squares signal the position of those stars considered to
be members of Cen\,OB1 only by Corti \& Orellana (in preparation).}
\label{fig:cuatro}
\end{figure*}

In the ideal case of a group of massive stars being at rest with respect to
 its local {\rm ISM}, it is expected that this group will be located at the
center of an hypothetical stellar wind bubble blown by this group of massive
stars. In the case the stars were not at rest with respect to its surroundings,
 the 
stars will likely occupied an eccentric location with respect to the bubble's
 centre. A look at Fig. \ref{fig:cuatro}, clearly shows that most of the Cen\,OB1 
stars are located towards the low latitude and low longitude sector of
{\rm GS}305+04-26. In  order to attempt to explain, in general terms, this
location, we shall follow the procedure outlined by 
\citet{slu03} (after correcting the sign of the w$_\odot$ term in their
 equation 19) and \citet{cic08}.

 Using the mean proper motions of Cen\,OB1 
(($\mu_{\alpha}cos{\delta}$, $\mu_{\delta}$)=(-4.65$\pm$0.15, -0.92$\pm$0.12).), 
a radial velocity of $-$26$\pm$1 \kms and a distance of 2.5$\pm$0.9  kpc 
(radial velocity and distance of {\rm GS}305+04-26) as the 
original distance of Cen\,OB1, the stellar peculiar velocity components are

v$_{pec,r}$ = $-$2.2 $\pm$ 14.8 \kms,

v$_{pec,l}$ = 12.4 $\pm$ 30.8 \kms,

v$_{pec,b}$ = $-$15.6 $\pm$ 4.1 \kms,

\noindent where v$_{pec,r}$,  is the peculiar radial velocity, v$_{pec,l}$ 
the peculiar
velocity along galactic longitude, and v$_{pec,b}$  the peculiar velocity 
along galactic latitude. The corresponding {\it spatial} peculiar velocity of 
the star, 

\begin{equation}
 v_{pec}= \sqrt{v^2_{pec,r}  + v^2_{pec,l} + v^2_{pec,b}},
\end{equation}

\noindent is v$_{pec}$= 20.0 $\pm$ 18.0 \kms. 

According to \citet{slu03} only two angles, labelled {\it i} and {\it $\phi$},
 are needed to completely define
the spatial orientation with respect to the observer of the structure created
by a star moving with respect to its {\rm ISM} with a high peculiar velocity. 
The angle {\it i} is measured from the line of sight towards the observer
 ({\it i}= 90\degr corresponds to the plane of the sky), whilst $\phi$ is 
measured on the plane of the sky,  counterclockwise from the tip of the
 peculiar velocity vector (see Fig. 4 of \citet{slu03}). Following 
\citet{slu03}, the above angles are defined as

\begin{equation}
 i=\arcsin (\frac{v_{pec,r}}{v_{pec}})
\end{equation}
\begin{equation}
\phi=\arctan (\frac{v_{pec,l}}{v_{pec,b}}) + \varepsilon
\end{equation}

where $\varepsilon$ is the angle between the line of constant declination 
and the line of constant galactic latitude at the position of the star.
In our case {$\varepsilon\sim$ 1\degr, and} the angles are 
{\it i} = 95\degr $\pm$ 45\degr, and {\it $\phi$} = 140\degr $\pm$70\degr.

At first glance, one may think that all stars of Cen\,OB1 should be
seen projected onto the interior of  {\rm GS}305+04-26. Nonetheless, a look at
 Fig. \ref{fig:cuatro}, taking into account, the uncertainties of the major 
 and minor semi-axes of  {\rm GS}305+04-26 (error bars 
drawn in the lower left corner), shows that  12 stars are definitively located
outside the large H{\sc i} shell. Bearing this in mind, under the assumption
that these peculiar velocities are inherent to the process of their formation,
the sequence of events put forward by \citet{sanc74} could explain the external position of the
 stars of Cen\,OB1 with respect to the expanding H{\sc i} shell. Certainly, in 
the case of Cen\,OB1 stellar winds and not SN$_e$, as was the case in Per\,OB2,
is the main triggering mechanism of the expansion of {\rm GS}305+04-26. 
Though these stars have
all the basic ingredients to create around them their own {\rm IB} (that
should be observable as a minimum in the H{\sc i} brightness temperature 
distribution towards these stars), these minima are not observed. The lack of
 detection may be explained by the following facts: {\it i)} the stars 
in question are seen projected onto the galactic plane where the H{\sc i} 
 brightness temperature is high and varies quite rapidly with position; 
 {\it ii)} The dual-value radial velocity
relationship implies that any H{\sc i} minimum related to an {\rm IB} 
located at the near kinematic distance will
very likely be filled in by H{\sc i} emission arising from the far kinematic 
distance region. Based on above, {\rm IB} around individual massive stars will
 be hard to recognize as such when the stars are seen projected towards the
 galactic plane and the 21-cm line emission is observed with a 
radiotelescope like the one used to produce the {\rm LAB} H{\sc i} all-sky
survey.

Summing up, and within their large uncertainties, the bulk peculiar
 spatial
velocity of the stars of Cen\,OB1 and the angles {\it i} and $\phi$ are
consistent with the present day location of Cen\,OB1 with respect to the
 centroid of {\rm GS}305+04-26.The off-center location of PSR J1253-5820 and 
PSR J1254-6150 is quite consistent with the peculiar spatial velocities of 
these objects \citep{cord98}.

\begin{figure*}
\centering
\includegraphics[width=20cm]{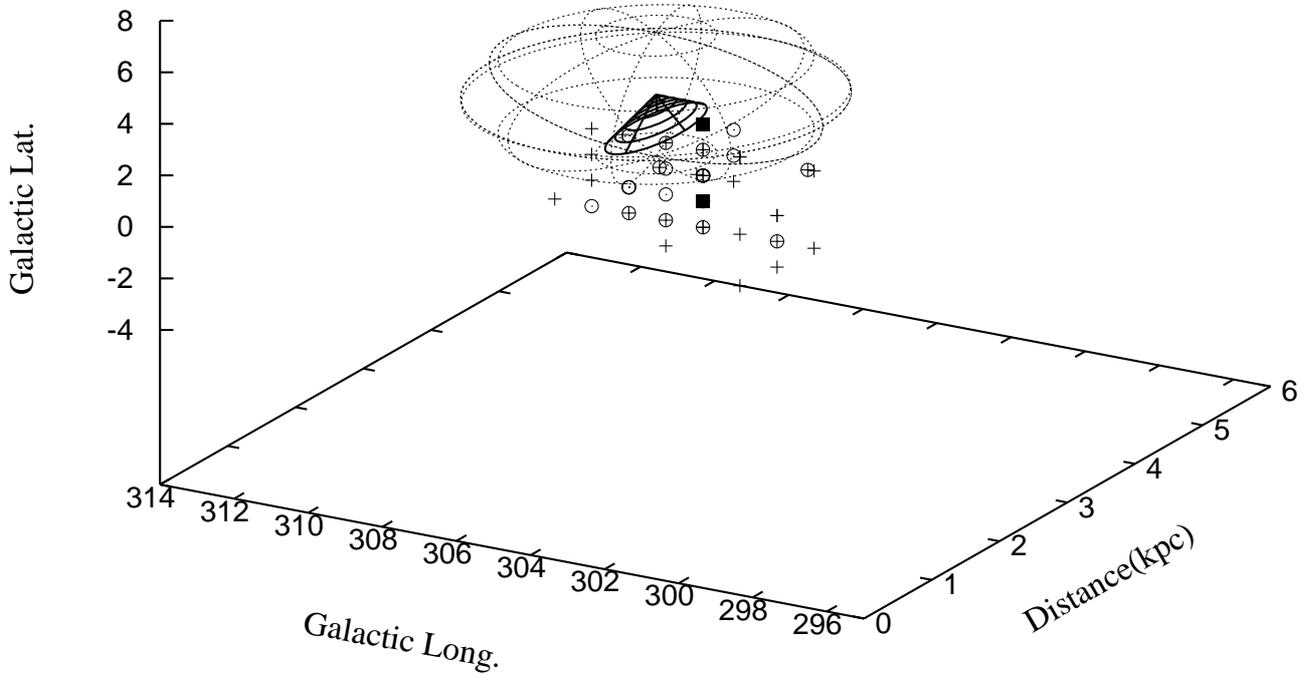}
\caption{Three dimensional sketch of {\rm GS}305+04-26 and the stars members of
 Cen\,OB1. The large H{\sc i} shell is shown as a dotted ellipsoid.
The location of both pulsars is given by fill-in black squares. The open
circles indicate the location of Cen\,OB1 members identified
as such by both \citet{hum84} and Corti \& Orellana (in preparation) whilst
the plus sign mark the position of those stars considered to
be members of Cen\,OB1 only by Corti \& Orellana (in preparation). 
The projection of the conical structure refered to in the text is also 
depicted in this Figure.}
\label{fig:cinco}
\end{figure*}

To explain the location of the OB-association projected onto the
southermost border of {\rm GS}305+04-26, a peculiar velocity of 10 $\pm$ 5 \kms
 would be needed along 12.6 $\pm$ 5 Myr (the age of the association). This
velocity is, within its large error, consistent with the bulk peculiar
motions derived for the stars of Cen\,OB1. A three dimensional sketch of the 
relative location of the stars with respect to {\rm GS}305+04-26, is given
in Fig. \ref{fig:cinco}. There, the supershell is represented by the ellipsoid
 shown in dotted lines. The large triangle-like symbol depicts the projection
 of a conical structure whose apex is located at the centre of 
{\rm GS}305+04-26, and its bisecting line is oriented along the angles {\it i} 
and $\phi$ derived above. The base of the cone points towards both lower
galactic latitudes and longitudes. The aperture of the cone is mostly given by 
the uncertainty in determining the angle $\phi$.

\section{Conclusions}

Examining the neutral hydrogen distribution towards the area covered by 
Cen\,OB1, in the light of new data that redefine the stellar membership of this
association, a new large H{\sc i} shell has been found. This structure,
designated {\rm GS}305+04-26, is elliptical in shape with major and minor
axe of 440 and 270 pc, respectively. The velocity interval where {\rm GS}305+04-26
is observable spans the range from $\sim$ -33 to $\sim$ -17 \kms. The
central velocity is -26$\pm$1 \kms and its expansion velocity is about
$\sim$8 \kms. The kinematic distance of this large structures is
 2.5$\pm$0.9 kpc, and has a total gaseous mass of about 
M$_t$ = (2.4$\pm$0.5) $\times$ 10$^5$ M$_{\odot}$.

Evidence has been provided in the sense that the stellar winds of the 54 stars
likely to be members of Cen\,OB1, may well explained the kinetic energy
of this large structure. The pulsars PSR J1253-5820 and PSR J1254-6150 have
both distances and characteristic ages compatible with the idea that they were 
born as the results of {\rm SN}$_e$ undergone in the past by massive members of
 Cen\,OB1.

The present day eccentric location of the stars of Cen\,OB1 with respect to the
centroid of {\rm GS}305+04-26 can be understood by the peculiar velocities of
the stars with respect to its local {\rm ISM}. 

Comparing the proper motion of the Wolf-Rayet star {\rm WR}\,48 with the bulk
 proper motions of
Cen\,OB1, it is highly unlikely that the Wolf-Rayet star can be considered a 
member of the stellar association.

\begin{acknowledgements}
This work was partially supported  by Consejo Nacional de Investigaciones 
Cient\'{\i}ficas y T\'ecnicas {\rm CONICET} under projects {\rm PIP} 
01299 and 01359, and by Universidad Nacional de La Plata ({\rm UNLP})
 under projects 11/G091 and 11/G096.
One of us (M.A. Corti) would like to thank both Dra E Gularte Scarone for her
 help in preparing Fig. \ref{fig:cinco} and Dr. N. Walborn for providing us
with updated information of some of the stellar data quoted in Table 
\ref{table:two}.
We would like to thank the referee for her/his constructive remarks.
\end{acknowledgements}

\bibliographystyle{aa}  
\bibliography{biblio170412}
   
\newpage

\newpage
\onecolumn
\begin{table*}[h!]
\begin{center}
\leavevmode
\caption{Information about the possible members of Centaurus OB1}
\label{table:two}
\begin{tabular}{|c|c|r|l|c|l|c|r|r|c|}
\hline
 ID &  lg &  bg	& SpT &	V & RV(LSR) & d & M &  t & Comments  \\
   &   ($\circ$)  &  ($\circ$) & &  (mag.) & (km s$^{-1}$) &  (kpc) & (M$_{\odot}$) & (10$^6$ yrs) &\\
\hline 

HD 108516  &	300.3 & -0.2 &	B2 III & 9.5 & &  2.8 $\pm$ 0.6 & 12.0  &	19.0 & \\
HD 109150  &	300.6 &	3.3 &  	B2 II-III & 8.6 & &  3.0 $\pm$ 0.7 & 14.5 & 14.0 & \\	
HD 109253 &	301.0 &	-1.6 &	B8 II-III c & 9.6 & &  2.3 $\pm$ 0.5  & -- & --	 & \\
HD 109505 &	301.1 &	1.2 &	B2 II c & 8.0 & &  2.1 $\pm$ 0.5 & 17.0 & 11.0 & \\
HD 109937 & 301.6 & -0.5 & B2-3 III & 9.4 & & 2.0 $\pm$ 0.5 & 10.5 & 24.0 & \\
HD 109978 &	301.6 &	0.0 &	O9 IV &	8.9 & -13   $\pm$ 5$^{(3)}$  & 2.7 $\pm$ 0.6 & 22.0 & 8.1 & \\
HD 110449 &	301.9 &	1.9 &	B2 III c & 9.3 & & 2.6 $\pm$ 0.6 & 12.0 & 19.0 & \\
HD 110597 &	302.0 &	3.5 &	B8 II &	9.5 & & 2.3 $\pm$ 0.5 & 11.4 & 21.0 & \\
CD-64 654 &	302.3 &	-2.7 &	B1 V &  9.7 & & 2.2 $\pm$ 0.5 & 13.4 & 16.0 & \\
HD 110912 &	302.3 &	2.1 &	B8 II-III & 9.8 & &  2.2 $\pm$ 0.5  & -- & -- & \\
HD 110984 &	302.4 &	1.7 &	B1 II-III & 9.0 & & 2.1 $\pm$ 0.5 & 20.0 & 9.2 & \\	
HD 111121 &	302.5 &	3.8 &	B8-9 II-III & 9.6 & &  2.7 $\pm$ 0.6  &	-- & -- & \\
HD 111579 &	302.8 &	1.6 &	B2 Ib-II & 9.1 & & 2.1 $\pm$ 0.5 & 19.5 & 9.5 &\\
CP-59 4552 & 303.2 & 2.5 & B1 III & 8.2 & -21$^{(3)}$ & 2.0 $\pm$ 0.5 & 15.4 & 13.0 & NGC 4755$^{(2)}$\\
HD 111904 &	303.2 &	2.5 &	B9 Ia$^{(1)}$   & 5.8 & -25  $\pm$ 3$^{(1)}$  & 2.6 $\pm$ 0.6 & 16.6 & 12.0 & NGC 4755$^{(2)}$ \\
HD 111934 &	303.2 &	2.5 &	B2 Ib$^{(1)}$   & 6.9 & -31  $\pm$ 2$^{(1)}$ & 2.0 $\pm$ 0.5 & 22.0 & 8.1 & NGC 4755$^{(2)}$  \\
HD 111973 &	303.2 &	2.5 &	B5 Ia$^{(1)}$   & 5.9 & -18 $\pm$ 4$^{(1)}$  &  2.7 $\pm$ 0.6 & 20.0 & 9.2 & NGC 4755$^{(2)}$  \\
HD 112026 &	303.3 &	2.0 &	B0.5 IV & 8.7 &  & 2.7 $\pm$ 0.6 & 16.3 & 12.0 & \\
HD 112147 &	303.4 &	3.8 &	B0 IV pe & 9.1 & & 3.0 $\pm$ 0.7 &  18.7 & 10.0 & \\
HD 112181 &	303.4 &	2.2 &	B2-3 II-III  &	8.8 &  & 2.0 $\pm$ 0.5 & 13.4 & 16.0 & \\
HD 112364 &	303.6 & 3.1 &   B0.5 Ib-a$^{(1)}$   & 7.4 & -68 $\pm$ 12$^{(1)}$  & 2.8 $\pm$ 0.6  & 25.0 & 7.0 & SB2 \\
HD 112366 & 303.6 & -0.6 & B6 Iab-b & 7.6 & -7  $\pm$ 2$^{(3)}$ & 2.1 $\pm$ 0.5 & 19.0 & 9.8 & \\ 
HD 112382 &	303.6 &	3.8 &	B4 II &	9.0 & & 2.6 $\pm$ 0.6 &15.0 & 14.0 & \\
HD 112471 &	303.7 &	2.0 &	B1 II-III & 8.8 & & 2.4 $\pm$ 0.6 & 20.0 & 9.2 & \\
HD 112485 &	303.7 &	2.0 &	B2 III &9.5 & -15$^{(3)}$ & 3.0 $\pm$ 0.7 & 12.0 & 19.0 & \\	
HD 112497 &	303.7 &	1.8 &	B1-2 II-III & 8.4 & &  2.2 $\pm$ 0.5  & 15.9 &	12.0 & \\
HD 112637 &	303.8 &	-0.4 &	B2 III c & 9.6 & & 2.2 $\pm$ 0.5 & 12.0 & 19.0 &\\
HD 112661 & 303.8 & 0.6 & B0.5 III-IV & 9.3 &  & 2.0 $\pm$ 0.5 & 17.0 & 11.3.0 & \\
HD 112784 &	304.0 &	2.3  &	O9.5 III & 8.3 & -34 $\pm$ 5$^{(3)}$  &  3.2 $\pm$ 0.7 & 21.7 & 8.3 & \\	
HD 112842 &	304.1 &	2.5 &	B3 Ib$^{(1)}$   & 7.1 & -33  $\pm$ 3$^{(1)}$  & 2.1 $\pm$ 0.5 & 21.1 & 8.6 & \\
HD 112852 &	304.1 &	3.3 &	B8-9 II &9.9 & & 2.6 $\pm$ 0.6 & 11.0 & 22.0 & \\
HD 113421 &	304.6 &	3 &	B0 Vn &	9.4 & -33$^{(3)}$  &  2.6 $\pm$ 0.6 &  17.5 & 11.0 & \\
HD 113422 &	304.5 &	1.1 &	B1 Ia$^{(1)}$   & 8.3 & -57  $\pm$ 4$^{(1)}$  & 3.2 $\pm$ 0.7 & 23.0 & 7.7 & \\
HD 113432 &	304.4 &	-0.7 & B1 Ib$^{(1)}$   & 8.9 & -21$\pm$11$^{(1)}$  & 2.7 $\pm$ 0.6 & 23.0 & 7.7 & \\
HD 113606 &	304.5 &	-1.4 &	O7 & 8.7 & & 2.3 $\pm$ 0.5 & 31.4 & 5.6 & \\
HD 113742 &	304.7 &	0.9 &	B1-2 III &9.2 & &  2.4 $\pm$ 0.5  & 13.7 & 16.0 & \\
HD 113754 &	304.7 &	-0.2 & O6-7 &	9.5 & -25 $\pm$ 2$^{(3)}$ &  2.8 $\pm$ 0.6 & 34.2 & 5.2 & \\	
CP-61 3452 &	304.8 &	1.0 &	O9.5 II c & 9.3 & & 2.8 $\pm$ 0.6 & 22.0 & 8.1 & \\
HD 114011 &	305.0 &	1.6 &	B4 Ib-II & 9.3 & -34 $\pm$ 5$^{(3)}$ & 3.1 $\pm$ 0.7  & 17.7 & 11.0 & \\	
HD 114213 &   305.2 &  1.3 &  B1 Ia-b$^{(1)}$   & 8.9 & -2 $\pm$ 6$^{(1)}$  &  2.2 $\pm$ 0.5 & 23.0 &  7.7 & \\
HD 114394 &	305.4 &	3.2 &	B0.5 IV (N) &	8.2 & & 2.1 $\pm$ 0.5 & 16.3 & 12.0 &\\
HD 114478 &	305.3 &	0.0 &	B1 Ib-II & 8.7 & -1$^{(3)}$ &  2.8 $\pm$ 0.6 & 21.1 & 8.6 & \\
HD 114669 &	305.4 &	0.2 &	B5 II &	9.1 & & 2.7 $\pm$ 0.6 & 14.5 & 14.0 & \\
HD 114737 &	305.4 &	0.1 &	O9 III &8.0 & & 2.8 $\pm$ 0.6 & 23.4 & 7.6 &\\
CP-60 4528 &	305.5 &	1.7 &	B1 III &8.8 & -25$^{(3)}$ & 2.2 $\pm$ 0.5 &  15.4 & 13.0 & \\
CP-60 4551 &	305.8 &	1.8 &	B1 III (N)e &	9.8 & & 2.3 $\pm$ 0.5 & 15.4 & 13.0 & \\
HD 115223 &	306.4 &	3.7 &	A0 Ib-II & 8.7 & & 3.0 $\pm$ 0.7 & 13.0 & 17.0 & \\	
HD 115455 &	306.1 &	0 &	O7.5 III &8.0 & -44 $\pm$ 14$^{(3)}$ & 3.1 $\pm$ 0.7 & 29.0 & 6.0 & \\	
HD 115497 &	306.4 &	3.7 &	B8-9 II&9.7 &  & 2.8 $\pm$ 0.6 & 11.0 & 22.0 & \\	
CP-61 3576 &	306.1 &	0.2 &	B0.5 V &9.5 & -40   $\pm$ 69$^{(3)}$ & 2.2 $\pm$ 0.5 & 14.9 & 13.7 & \\
HD 115666 &	306.4 &	1.8 &	B7 IIc &9.4 &  & 2.8 $\pm$ 0.6 & 12.5 & 18.0 & \\	
HD 116121 &  306.5  & -0.4 & O9.5 V & 9.3 & & 3.4 $\pm$ 0.8 & 19.0 & 9.8 & \\ 
HD 116403 &	307.3 &	1.6 &	B5 II-III & 9.0 & & 3.1 $\pm$ 0.7 & 10.7 & 23.0 &\\ 
HD 116864 &	307.0 &	-0.6 &	B9 II-III & 9.1 & & 2.4 $\pm$ 0.5  & -- & -- & \\
\hline 
\end{tabular}
\end{center}
\end{table*}
 
\noindent \parbox{14cm}{{\scriptsize $^{(1)}$ (Corti \& Orellana, in preparation)}}\\
\parbox{14cm}{{\scriptsize $^{(2)}$ \citet{hum78}}}\\
\parbox{14cm}{{\scriptsize $^{(3)}$ Obtained of the SIMBAD Astronomical Database ({\rm CDS}).}}\\ 

\IfFileExists{\jobname.bbl}{}
{\typeout{}
\typeout{****************************************************}
\typeout{****************************************************}
\typeout{** the bibliography and then re-run LaTeX}
\typeout{** twice to fix the references!}
\typeout{****************************************************}
\typeout{****************************************************}
\typeout{}
}
\end{document}